 \documentclass[aps,prl,twocolumn,superscriptaddress,amsmath,amssymb,showpacs]{revtex4}
\usepackage{graphicx}
\usepackage{amsmath}
\usepackage{amssymb}
\usepackage{mathrsfs}
\usepackage{amsfonts}
\usepackage{dcolumn}
\usepackage{bm}

\providecommand{\abs}[1]{\left\lvert#1\right\rvert}

\newcommand{\lbar}{{\lambda \negthickspace \! \text{\rule[5.5pt]{0.135cm}{.2pt}} \, }}
\begin{document}
\title{Duality Between Spatial and Angular Shift in Optical Reflection}

\author{A.~Aiello}
\email{Andrea.Aiello@mpl.mpg.de}
\affiliation{Max Planck Institute for the Science of Light, G\"{u}nter-Scharowsky-Stra{\ss}e 1/Bau 24, 91058 Erlangen, Germany}
\affiliation{Huygens Laboratory, Leiden University,
P.O.\ Box 9504, 2300 RA Leiden, The Netherlands}
\author{M.~Merano}
\affiliation{Huygens Laboratory, Leiden University,
P.O.\ Box 9504, 2300 RA Leiden, The Netherlands}
\author{J.~P.~Woerdman}
\affiliation{Huygens Laboratory, Leiden University,
P.O.\ Box 9504, 2300 RA Leiden, The Netherlands}
\begin{abstract}
We report a unified representation of the spatial and angular Goos-H\"{a}nchen  and Imbert-Fedorov shifts that occur when a light beam reflects from a plane interface. We thus reveal the dual nature of spatial and angular shifts in optical beam reflection. In the  Goos-H\"{a}nchen case we show  theoretically and experimentally that this unification naturally arises in the context of reflection from a lossy surface (e.g., a metal).
\end{abstract}
 \maketitle
%
%
%
%
%
%
%
%
%
{\flushleft\emph{\hspace{.3cm}Introduction.$\,$}\rule[1.8pt]{0.3cm}{.4pt}}
In the $17$th century  Newton  was the first to surmise that the center of a reflected beam should present  a small spatial shift $\Delta$ in the plane of incidence, relative to its geometrical optics position \cite{NewtonOptiks}.
 More than two centuries afterwards, in $1947$ Goos and H\"{a}nchen (GH) \cite{GH} were able to quantitatively measure such a  shift  (see \cite{NatPhoton.3.337} for a literature survey since that time). The GH shift is typically in the sub-wavelength domain; it has become technologically important in recent years since it directly affects the modes of optical waveguides and microcavities \cite{Foster:07,unterhinninghofen:016201} and has great potential for (bio)sensor applications \cite{yin:261108}. Theoretically, the GH shift has been explained at various levels  and several generalizations have been discovered \cite{Porras199613,Nasalski:88,McGandC}.  Amongst the latter
 it was predicted that the axis of the reflected beam should display a small angular deviation $\Theta$ from the law of specular reflection $\theta_\text{inc}=\theta_\text{ref}$  \cite{Bertoni,Porras199613}.
Interestingly, it took about  $50$ years since the original experiment performed  by Goos and H\"{a}nchen  to actually observe such angular shift in the microwave \cite{Nimtz} and the optical \cite{NatPhoton.3.337} regimes.
Presently, it is common wisdom that spatial and angular GH shifts are two different phenomena observables in two mutually exclusive regimes: the spatial GH shift occurs  in total reflection (reflected intensity $=$ incident intensity)  \cite{Artmann}, while the angular GH shift  appears in partial reflection (reflected intensity $<$ incident intensity) \cite{Bertoni}.

In this Letter we show that this separation is artificial.  We present a unified description for the spatial and angular GH shifts that will appear as two aspects of a unique beam-propagation phenomenon.
 We show that such  duality between spatial and angular shift is rather general and also
applies to the Imbert-Fedorov (IF) effect, which is a shift \emph{normal} to the plane of incidence \cite{CandI} that has drawn considerable interest lately \cite{OnodaEtAlPRL,BliokhPRL,HostenandKwiat}. Finally,  for the GH shift we show that
 unification naturally arises in the context  of reflection  from lossy surfaces.
 For this case we also furnish an experimental demonstration that the spatial and angular shifts  occur \emph{simultaneously}.

Our Letter is structured as follows: We give first a qualitative picture of the envisaged unification. Then, we furnish a rigorous theoretical analysis of the beam-propagation problem and show that the unified description actually holds for both the GH and the IF shifts. Finally, we demonstrate, both theoretically and experimentally, that for the GH case lossy reflecting surfaces naturally induce unification.
{\flushleft\emph{\hspace{.3cm}Qualitative picture.$\,$}\rule[1.8pt]{0.3cm}{.4pt}}
%
%
%
%
\begin{figure}[!!br]
\includegraphics[angle=0,width=7.5truecm]{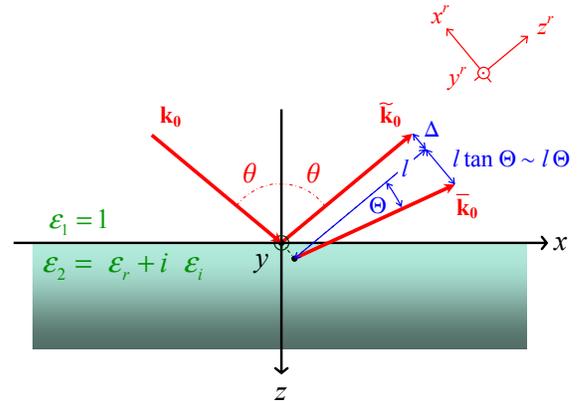}
\caption{\label{fig:1} (color online).
Scheme of the beam reflection at the plane interface.  Here  $\widetilde{\mathbf{k}}_0$ and $\overline{\mathbf{k}}_0$ are the central wave vectors of the reflected beam as predicted by geometrical and wave optics, respectively. The waist of the incident beam is located at the origin of the laboratory Cartesian frame $xyz$.}
\end{figure}
%
%
%
Consider a system consisting of two homogeneous, isotropic media of dielectric constants $\varepsilon_1$ and $\varepsilon_2$, filling the half-spaces $z<0$ and $z \geq 0$, respectively, as shown in Fig. 1.
A monochromatic beam of light of wavelength $\lambda_0$ and waist $w_0$  propagates along the central wave vector ${\mathbf{k}}_0$ in the region $z<0$ before impinging upon the  plane interface of equation $z=0$ that separates medium $1$ from medium $2$.
Detailed derivations of the angular and  the spatial displacements $\Theta$  and $\Delta$ for this system have already been reported elsewhere \cite{Artmann,McGandC,Nasalski:88,Porras199613,AielloOL2008,AielloOL2009,AielloGH2009} and will not be repeated here. We merely quote the basic results in the form
\begin{align}\label{GHs&a}
\Delta_\lambda =  \lbar_0 \, \text{Im} \left[ D_\lambda \right], \qquad \Theta_\lambda =  - ({\theta_0^2}/{2})\text{Re} \left[ D_\lambda \right],
\end{align}
where $\lbar_0 = \lambda_0/(2 \pi)$, and $\theta_0 = 2 \lbar_0/w_0$ is the angular spread of the incident beam \cite{MandelBook}. The expressions above are valid for both the GH and the IF shifts where the coefficient $D_\lambda$ is equal to
\begin{align}\label{D}
D_\lambda =   \frac{\partial \ln r_\lambda}{\partial \theta} =\frac{1}{R_\lambda}\frac{\partial R_\lambda}{\partial \theta} + i    \frac{\partial \phi_\lambda}{\partial \theta} \equiv D_\lambda^\mathrm{R} + i D_\lambda^\mathrm{I},
\end{align}
in the GH case,  and to $D_\lambda =   2 \, i  [(r_P + r_S)/{r_\lambda}] \cot \theta$
%
%
%
in the IF case \cite{NoteIF2}.
Here $D_\lambda^\mathrm{R}  \equiv \text{Re}[D_\lambda]$,  $D_\lambda^\mathrm{I}  \equiv \text{Im}[D_\lambda]$, and the index $\lambda$ is a label for the two linear polarizations parallel ($\lambda =P$ or \textsl{TM}) and perpendicular ($\lambda =S$ or \textsl{TE}) to the central plane of incidence $x\text{-}z$. Moreover,
$r_\lambda \equiv r_\lambda(\theta) = R_\lambda \exp(i \phi_\lambda)$ is the Fresnel reflection coefficient \cite{BandWBook} evaluated at the central angle of incidence $\theta$, where $ R_\lambda = \abs{r_\lambda}$ and $\phi_\lambda = \arg{r_\lambda}$.

From Fig. \ref{fig:1} it follows, using elementary geometrical considerations,  that the total beam displacement observed at distance $l$ from the origin is expressible as a \emph{linear combination} of  $\Delta_\lambda$ and $\Theta_\lambda $ (supposedly $\Theta_\lambda \ll 1$) of the form:
\begin{align}\label{Total}
\delta_\lambda(l) =   \Delta_\lambda + l \, \Theta_\lambda \, .
\end{align}
From now on, in order to avoid trivial repetitions, we will consider explicitly the GH case only. However, all our conclusions for the GH case can be straightforwardly extended to the IF case with minimum effort.

Equation (\ref{Total})  supports the hypothesis that $\Delta_\lambda$ and $\Theta_\lambda$ are two different manifestations  of a unique phenomenon, as they can be connected by such an elementary geometric relation.
From Eq. (\ref{D}) it follows that in the case of total reflection (${R_\lambda}=1$),
 the reflection coefficients reduce to a pure phase factor  $r_\lambda =\exp(i \phi_\lambda)$, thus $D_\lambda^\mathrm{R} =0$ and the reflected beam undergoes a purely spatial shift $\Delta_\lambda$. Vice versa, if the reflection coefficients are strictly real, as in the case of air-to-glass partial reflection, then $\phi_\lambda \in \{ 0,\pm \pi \} \, \Rightarrow \,D_\lambda^\mathrm{I} =0$, and the beam undergoes a purely angular deflection $\Theta_\lambda$.
However, when the reflection coefficients are \emph{complex}, then both  $\Delta_\lambda \neq 0$ and $\Theta_\lambda \neq 0$, and a unified description of spatial and angular GH shifts becomes mandatory.
Since for a plane interface between two lossless media the Fresnel coefficients are either purely real or pure phase factors \cite{BandWBook}, then we have  either  $\delta_\lambda =  l \, \Theta_\lambda$ or $\delta_\lambda =   \Delta_\lambda$, respectively, and spatial and angular GH shifts are mutually exclusive.
On the other hand, when either  one or both of the media are lossy, the separation between spatial and angular GH shifts becomes artificial, and unification naturally occurs.
{\flushleft\emph{\hspace{.3cm}Theoretical description.$\,$}\rule[1.8pt]{0.3cm}{.4pt}}
As a first step towards unification, in this section we aim to deduce Eq. (\ref{Total}) from a perfectly general \emph{ab initio} calculation, without restrictions  on the media forming the interface or on the transverse shape of the light beam.

It is well known that under minimal hypotheses \cite{McGandC,MandelBook} it is possible to obtain a valid angular spectrum representation for the electric field of the reflected beam of the form $\mathbf{A}(\mathbf{X},Z)
 = ({2 \pi})^{-1}\iint\, \mathbf{A}(\bm{\kappa},Z) \, e^{i\bm{\kappa}\cdot \mathbf{X}}\,{ d^{\,2} \kappa }$, where $\mathbf{A}(\bm{\kappa},Z)$ is uniquely determined by the angular spectrum of the incident beam and the Fresnel reflection coefficients \cite{AielloGH2009}. A Cartesian reference frame attached to the reflected beam with a scaled coordinate system is utilized in which $X = k_0 x^r, \, Y = k_0 y^r, \, Z = k_0 z^r$, and $k_0=\abs{\mathbf{k}_0}$, as shown in Fig. (\ref{fig:1}). Moreover, we have defined   $\mathbf{X} \equiv X \hat{\bm{x}}^r + Y \hat{\bm{y}}^r$, and
$\bm{\kappa}=  \mathbf{k} - \, \hat{\bm{z}}^r(\hat{\bm{z}}^r \cdot \mathbf{k} ) \equiv \kappa_1 \hat{\bm{x}}^r + \kappa_2 \hat{\bm{y}}^r$ is the transverse part of the unit wave vector $\hat{\mathbf{k}} \equiv \mathbf{k}/k_0 = \bm{\kappa} + \kappa_3  \hat{\bm{z}}^r$ with respect
to $\hat{\bm{z}}^r \equiv \widetilde{\mathbf{k}}_0/k_0$, with $\kappa_3 =(1 - \bm{\kappa} \cdot \bm{\kappa})^{1/2}$. Here   $\widetilde{\mathbf{k}}_0 =  \mathbf{k}_0 - 2 \, \hat{\bm{z}}(\hat{\bm{z}} \cdot \mathbf{k}_0 )$ is the central wave vector of the reflected beam as ruled by geometrical optics. The angular spectrum in the plane $Z$ is determined by its value at $Z=0$ via the relation \cite{MandelBook}:
\begin{align}\label{ArefZ}
\mathbf{A}(\bm{\kappa},Z) = \mathbf{A}(\bm{\kappa},0)\exp(-i \mathcal{H} Z),
\end{align}
where $\mathcal{H} = -\kappa_3 $ is the so-called \emph{optical Hamiltonian} that governs the well-known Hamilton equations of motion for light rays in vacuum  \cite{sekiguchi:830}:
\begin{align}\label{Heqn}
\frac{d \bm{\kappa}}{d Z} = - \frac{ \partial \mathcal{H}}{\partial \mathbf{X}} = \bm{0}, \qquad \frac{d \mathbf{X}}{d Z} =  \frac{ \partial \mathcal{H}}{\partial  \bm{\kappa}} = \frac{ \bm{\kappa}}{\kappa_3}.
\end{align}
The analogy between Eq. (\ref{ArefZ}) and the expression for the time evolution  of the wave function of a quantum system  in the Schr\"{o}dinger picture,  suggests the use of an enlightening quantum-like notation \cite{Gloge:1629,Stoler1981,Berry03} by writing $\mathbf{A}(\bm{\kappa},Z) = \langle \bm{\kappa} | \mathbf{A}(Z) \rangle$ and  $\mathbf{A}(\mathbf{X},Z) = \langle \mathbf{X} | \mathbf{A}(Z) \rangle$,
where $| \bm{\kappa} \rangle = | \kappa_1, \kappa_2 \rangle$ and $| \mathbf{X} \rangle = | X, Y \rangle$ are the basis vectors in the transverse momentum and position space, respectively. In Eq. (\ref{ArefZ})  the longitudinal coordinate $Z$ has the role of a dimensionless time, then we can write $| \mathbf{A}(Z) \rangle = \exp(-i \hat{\mathcal{H}} Z)| \mathbf{A}(0) \rangle$, where $\hat{\mathcal{H}}$ is the Hamiltonian  operator defined via  $  \langle \bm{\kappa}  | \hat{\mathcal{H}}  | \bm{\kappa}'  \rangle
 =  - (1 - \bm{\kappa} \cdot \bm{\kappa})^{1/2}  \delta(\bm{\kappa} - \bm{\kappa}')$.
In quantum mechanics unitary evolution implies that  probabilities are conserved along with propagation, namely  $\langle \mathbf{A}(Z) | \mathbf{A}(Z)\rangle = \langle \mathbf{A}(0) | \mathbf{A}(0)\rangle$. In our case, this means that the flux of the electric field energy density across any plane $Z=\text{const.}$, is independent from $Z$, namely $\iint\, \bigl|\mathbf{A}(\mathbf{X},Z)\bigr|^2 d X d Y = \iint\, \bigl|\mathbf{A}(\bm{\kappa},0)\bigr|^2 \,d^{\,2} \kappa =1$,
where we have renormalized the  field amplitude of the reflected beam to ensure $\langle \mathbf{A}(0) | \mathbf{A}(0)\rangle =1$.
At any given coordinate $Z$ the electric field energy density $\bigl|\mathbf{A}(\mathbf{X},Z)\bigr|^2$ gives the spatial beam profile in the observation plane $X\text{-}Y$. The $Z$-dependent centroid of such energy distribution $\left\langle\mathbf{X} \right\rangle(Z)= \iint\, \mathbf{X} \bigl|\mathbf{A}(\mathbf{X},Z)\bigr|^2 d X d Y$,
measures the deviation of the beam axis with respect to the central axis $\hat{\bm{z}}^r$ \cite{AielloOL2008,Porras199613} defined by geometrical optics.
If we define the transverse position and momentum operators such that $\hat{\mathbf{X}}|\mathbf{X}' \rangle = \mathbf{X}'|\mathbf{X}' \rangle$ and $\hat{\mathbf{K}}|\bm{\kappa}' \rangle = \bm{\kappa}'|\bm{\kappa}' \rangle$, respectively, then the centroid of the beam can be evaluated as $\left\langle\mathbf{X} \right\rangle(Z) = \langle \mathbf{A}(Z)|\hat{\mathbf{X}} |  \mathbf{A}(Z) \rangle =$
$ \langle \mathbf{A}(0)|e^{i \hat{\mathcal{H}} Z}\hat{\mathbf{X}}e^{-i \hat{\mathcal{H}} Z} |  \mathbf{A}(0) \rangle = \langle \mathbf{A}(0)|\hat{\mathbf{X}}_H(Z) |  \mathbf{A}(0) \rangle$, where $\hat{\mathbf{X}}_H(Z)$ is the position operator in the Heisenberg picture:
\begin{align}\label{Xk}
\langle \bm{\kappa}| \hat{\mathbf{X}}_H(Z) | \bm{\kappa}' \rangle
 =  \left( -\frac{1}{i}\frac{\partial}{\partial \bm{\kappa}}  + Z  \frac{\bm{\kappa}}{\kappa_3} \right) \delta(\bm{\kappa} - \bm{\kappa}').
\end{align}
The first term on the right side of Eq. (\ref{Xk}) coincides with the momentum-representation of the $Z$-independent position operator $\hat{\mathbf{X}}$ in the Schr\"{o}dinger picture \cite{NoteQuantum}. Therefore, from the very definition of $\left\langle\mathbf{X} \right\rangle(Z)$,  it  follows that the expectation value $\langle \mathbf{A}(0)|\hat{\mathbf{X}} |  \mathbf{A}(0) \rangle = \left\langle\mathbf{X} \right\rangle(0)$ gives both the GH and IF  \emph{spatial} shifts $\langle {\hat{X}_H}(0) \rangle$ and $\langle {\hat{Y}_H(0)} \rangle$, respectively.
 The second term, which is linear in $Z$, is proportional to right side of
 the second equation in (\ref{Heqn}) which determines the direction of propagation of  classical rays of light, since  $\bm{\kappa}/\kappa_3 = \tan \vartheta_1 \hat{\bm{x}}^r + \tan \vartheta_2 \hat{\bm{y}}^r$. As we expect small angular deviations from geometrical optics predictions, we can write $\tan \vartheta_i \approx \vartheta_i \; (i =1,2)$, and the exact relation between wave and geometrical optics  established by Eqs.  (\ref{ArefZ}-\ref{Heqn}) allows us to identify  $  {\partial \bigl\langle \hat{X}_H(Z) \bigr\rangle }/{\partial Z}  \approx \vartheta_1$ with the  angular GH shift, and  ${\partial \bigl\langle \hat{Y}_H(Z) \bigr\rangle }/{\partial Z}  \approx \vartheta_2$ with the angular IF shift.
Such identification becomes even more clear by noting that  from $[\hat{\mathbf{K}}, \hat{\mathcal{H}}] =0$ and Eq. (\ref{Xk}) it follows that
\begin{align}
\left\langle\mathbf{X} \right\rangle(Z)
 & = \bigl\langle \hat{\mathbf{X}}  \bigr\rangle + Z \big\langle {\hat{\mathbf{K}}}{\big(1 - \hat{\mathbf{K}} \cdot \hat{\mathbf{K}}\big)^{-1/2} }\, \big\rangle ,
  \label{Xk2}
\end{align}
which reduces to $\left\langle\mathbf{X} \right\rangle(Z)\approx \bigl\langle \hat{\mathbf{X}}  \bigr\rangle + Z \big\langle \hat{\mathbf{K}}\bigr\rangle$ for  $\vartheta_i \ll 1$ \cite{NoteParax2}, and the angular brackets  indicate expectation values with respect to the state $|  \mathbf{A}(0) \rangle$.
Equation (\ref{Xk2}) has the same form $\left\langle\mathbf{X} \right\rangle(Z) = \bm{\Delta} + Z \bm{\Theta}$ as Eq. (\ref{Total}), where  $\bm{\Delta}$ depends on the position operator $\hat{\mathbf{X}}$, and $\bm{\Theta}$  on the momentum operator $\hat{\mathbf{K}}$ solely,
 thus defining unambiguously both a spatial  and an angular  vector shift of the beam equal to $\bm{\Delta} = \left\langle\mathbf{X} \right\rangle(0)$ and  $\bm{\Theta} = \partial \left\langle\mathbf{X} \right\rangle(Z)/ \partial Z$, respectively. Note that the  dependence of Eq. (\ref{Xk2}) on the characteristics of the reflecting surface and on the polarization of the incident beam  is contained in the form of the state $|  \mathbf{A}(0) \rangle$.

Equation (\ref{Xk2}) establishes the first part of the announced unification by reproducing Eq. (\ref{Total}) that was naively deduced on the ground of simple geometric reasoning. We emphasize that, because of its vector form,  Eq. (\ref{Xk2}) describes both GH and IF shifts.
The next step is to
 demonstrate that reflection from lossy surfaces  induces \emph{simultaneously} both spatial and angular GH shifts.
{\flushleft\emph{\hspace{.3cm}Loss-induced unification.$\,$}\rule[1.8pt]{0.3cm}{.4pt}}
Consider again the system shown in Fig. 1. Assuming air as medium $1$, we can write the dielectric constant of medium $2$ as $\varepsilon_2 = \varepsilon_r + i \varepsilon_i$. Then, we can distinguish that medium $2$ is a  dielectric, namely $\varepsilon_r >1$; or a metal with $\varepsilon_r<0$. For both cases, from Eq. (\ref{D}) and the well-known expressions for the Fresnel reflection coefficients \cite{BandWBook}, it follows that
\begin{widetext}
\begin{equation}
D_P  =   \frac{2 \sin \theta}{\sqrt{\left(\varepsilon_r - \sin^2 \theta \right) + i \varepsilon_i}}\frac{  \varepsilon_i^2 + \varepsilon_r\left( 1 - \varepsilon_r\right)  + i \varepsilon_i \left( 1 - 2 \varepsilon_r\right)  }{
  \left(\varepsilon_r - \sin^2 \theta \right)  + \left(\varepsilon_i^2 -\varepsilon_r^2 \right)\cos^2 \theta  + i \varepsilon_i \left( 1 - 2 \varepsilon_r \cos^2 \theta \right) }, \quad D_S  =  \frac{2 \sin \theta}{\sqrt{\left(\varepsilon_r - \sin^2 \theta \right) + i \varepsilon_i}} . \label{DP}
\end{equation}
\end{widetext}
To elucidate the role of the losses, we  expand Eq. (\ref{DP}) in a Taylor series around $\varepsilon_i =0$, and keep terms up to the first order in $\varepsilon_i$.
For sake of simplicity, from now on we will consider only the metal case for which the Taylor expansion furnishes
\begin{align}
D_P^\mathrm{R} =&  \varepsilon_i \frac{\sin \theta\left(1 - \varepsilon_r \right)^2}{\left(  \sin^2 \theta -\varepsilon_r \right)^{3/2}} \frac{2 \sin^4 \theta - \varepsilon_r \left( \sin^2 \theta + \varepsilon_r \cos^2 \theta \right)}{ \left( \sin^2 \theta -\varepsilon_r \right)+\varepsilon_r^2 \cos^2 \theta },  \label{DP3r} \\
D_P^\mathrm{I} =&    \frac{-2 \varepsilon_r \sin \theta  }{\sqrt{ \sin^2 \theta -\varepsilon_r}
\left(\sin^2 \theta - \varepsilon_r \cos^2 \theta  \right)}, \label{DP3i}
\end{align}
 for $P$  polarization, and $D_S^\mathrm{R}  =  \varepsilon_i {2  \sin \theta  }/{\left(  \sin^2 \theta -\varepsilon_r \right)^{3/2}}$, $D_S^\mathrm{I} =  {-2  \sin \theta  }/{\sqrt{\sin^2 \theta -\varepsilon_r }}$,
for  $S$ polarization. From the expressions above we can see that for an ideal lossless metal ($\varepsilon_i=0$) we have that $D_\lambda$ is purely imaginary and only the spatial shift occurs \cite{Merano07}.
However, when losses are ``turned on'' by letting $0 < \varepsilon_i \ll 1$, a first-order {real} part must be added to  $D_\lambda^\mathrm{I}$,  causing the \emph{simultaneous} existence of both spatial and angular shifts.
%
%
%
\begin{figure}[!th]
\includegraphics[angle=0,width=8truecm]{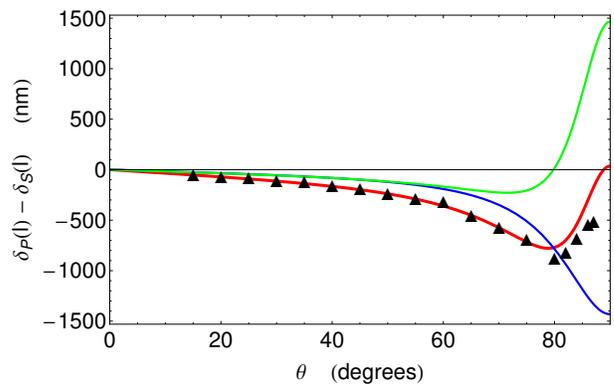}
\caption{\label{fig:2} (color online).
$\delta_P(l) - \delta_S(l)$  at $l = 11.9 \text{ cm}$ for a fundamental Gaussian beam with wavelength $\lambda_0 = 820 \; \text{nm}$, waist $w_0 = 32 \; \mu \text{m}$, incident on a gold surface with $\varepsilon_2 = -29.02 + i \, 2.03$. Blue line $=$  $\Delta_P-\Delta_S$; green line $=$   $l(\Theta_P-\Theta_S)$; red line $=$  $\delta_P - \delta_S$; black triangles $=$ measured  $\delta_P - \delta_S$. 
}
\end{figure}
%
%
Such loss-induced coexistence between $\Delta$ and $\Theta$ is clearly illustrated in Fig. 2 where Eq. (\ref{Total}) is displayed for  reflection at an air-gold interface. The red curve represents the total beam shift at distance $l$ from the origin and it is given, according to Eq. (\ref{Total}), by the sum of the green curve (angular shift), and the blue one (spatial shift).
This completes the second part of our unification program.
{\flushleft\emph{\hspace{.3cm}Experiment.$\,$}\rule[1.8pt]{0.3cm}{.4pt}}
The experimental set-up is sketched in Fig. 3.
%
%
 \begin{figure}
\includegraphics[angle=0,width=8truecm]{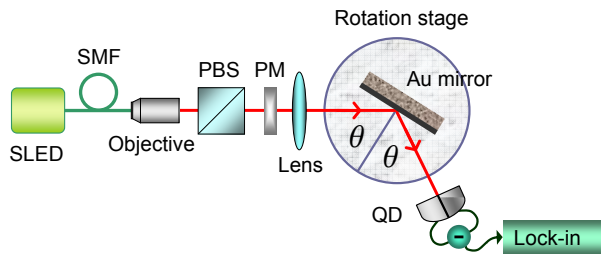}
\caption{\label{fig:3}  (color online). Layout of the experimental set-up. }
\end{figure}
%
%
%
The $820$-nm output of a
super-luminescent light-emitting diode (\textsf{SLED}) is spatially filtered by a single-mode
optical fiber (\textsf{SMF}) to select a \textsf{TEM}$_{00}$ mode. This is then
collimated by a microscope objective and sent through a Glan polarizing prism (\textsf{PBS}) to fix its polarization
to $P$.
 A polarization modulator (\textsf{PM}) switches the beam polarization between $P$ and $S$ to exploit the fact that both  spatial and  angular GH shifts are polarization dependent. Then, a lens is used to focus the beam to the desired spot size in front of the mirror. While the spatial GH shift is unaffected by such beam focusing, the angular GH shift depends on the beam angular aperture. The reason for this behavior is evident from the $\theta_0^2$ factor in Eq. (\ref{GHs&a}), and its physical origin is explained in  \cite{NatPhoton.3.337}.
Finally, the  difference-signal from a quadrant detector (\textsf{QD}) is fed into a lock-in amplifier in order to detect the beam displacement in the plane of incidence when polarization is switched between $S$ and $P$.  See \cite{NatPhoton.3.337,Merano09} for further details.

The results of the measurements are shown in Fig. \ref{fig:2} along with theoretical prediction. The good agreement confirms the simultaneous occurrence of the spatial and the angular GH shifts  in the lossy regime.
{\flushleft\emph{\hspace{.3cm}Conclusions.$\,$}\rule[1.8pt]{0.3cm}{.4pt}}
In this Letter we have presented a unified description for
 spatial and angular Goos-H\"{a}nchen shifts occurring in light beam reflection from lossy surfaces. Such description applies to the Imbert-Fedorov effect as well. The unification theory  has been worked out for a weakly-
absorbing metal and the corresponding GH shift has been observed experimentally in reflection from
an air-gold plane interface.


\begin{thebibliography}{30}
\bibitem[{Not({\natexlab{a}})}]{NoteIF2}
\bibinfo{note}{In this case we denote with $\Delta_\lambda$ and
  $\Theta_\lambda$, respectively, the spatial and angular separation between
  the two right-circularly and left-circularly polarized components of the
  reflected beam generated by the reflection-induced splitting of the
  $\lambda$-polarized ($\lambda \in \{P,S\}$) incident beam \cite{CandI}.}
\bibitem[{Not({\natexlab{b}})}]{NoteQuantum}
\bibinfo{note}{All the operators without the subscript ``$H$'' must be intended
  in the Schr\"{o}dinger picture.}
\bibitem[{Not({\natexlab{c}})}]{NoteParax2}
\bibinfo{note}{The reader knowledgeable about paraxial optics will easily
  recognize from Eq. (\ref{Xk2}) in the limit $\vartheta_i \ll 1$, and from the
  relation $\hat{\mathbf{K}}_H(Z) = \hat{\mathbf{K}}$, the free-space
  propagation $ABCD$-matrix law: $ \left( \begin{array}{c} \hat{\mathbf{X}}_H
  \\ \hat{\mathbf{K}}_H \end{array} \right)(Z) = \left( \begin{array}{cc} 1 & Z
  \\ 0 & 1 \\ \end{array} \right) \left( \begin{array}{c} \hat{\mathbf{X}} \\
  \hat{\mathbf{K}} \end{array} \right) $}.
\end{thebibliography}
\end{document}